\documentstyle[preprint,aps]{revtex}

\begin{document}  
\draft 

\title{Manipulating the superfluid - Mott insulator transition of a Bose -
Einstein condensate in an amplitude - modulated optical lattice.}

\author{G.M.Genkin$^*$.}

\address{ Department of Physics and Astronomy, Northwestern University,
Evanston, Illinois, 60208.}

\maketitle 
                 
\begin{abstract}  

The superfluid - Mott insulator transition in a BEC confined
in an amplitude - modulated  
optical lattice
can be manipulated by the modulation
strength.
Two standing laser waves of main and sideband frequencies of an optical lattice
induce a Raman transition;
  due to 
 resonant Raman 
driving the critical value of the transition parameter depends on
the modulation strength
 and the detuning from resonance
and can be tuned to a given value. It is shown that
there is an interval of the initial MI phase, in which
 a modulation switches the MI phase to the SF phase.

{*} Electronic address: gena@pluto.phys.nwu.edu.
\end{abstract}  

Manipulating a phase transition is a subject of great interest to physicists. A
physical system that crosses the boundary between two phases changes its
properties in a fundamental way. Second - order phase transitions usually
termed classical phase transitions, take place at a finite temperature, and
they are accomplished by changing the temperature. A quantum phase transition
is one that occurs at absolute zero of temperature. The quantum phase
transitions have attracted much interest in recent years. These transitions are
accomplished by changing not the temperature, but some parameter in the
Hamiltonian of the system. This parameter is the lattice laser intensity for a
Bose - Einstein condensate (BEC) in an optical lattice, which controls the
superfluid - Mott insulator transition; the charging energy in Josephson -
junction arrays. which controls the superconductor - insulator transition; the
magnetic field in a quantum - Hall system, which controls the transition
between the Hall plateaus; disorder in a conductor near its metal - insulator
transition. The competition between two interactions is fundamental to quantum
phase transitions and inherently different from classical phase transitions
which are driven by the competition between inner energy and entropy. The
quantum phase transition are driven by the interplay of the different
contributions to the Hamiltonian of the system. For instance, for a BEC in an
optical lattice the two terms in the Hamiltonian represent the self -
interaction of the atoms in the lattice site and the hopping of the atoms
between the sites, for Josephson - junction arrays they represent the charging
energy of each grain and the Josephson coupling energy between grains; the
competition between them determines the quantum phase transition.

The experimental observation of Bose-Einstein condensation in a dilute gas of
ultracold trapped atoms [ 1-3 ] has generated much interest in manipulating
such coherent matter by external electromagnetic fields. Recently atoms have
been confined in optical potentials created by standing light waves [ 4, 5 ],
with the wavelength of the optical potential much smaller than the dimensions
of the trap. The unique prospects of this new class of system (the correlated
bosons on a lattice [ 6 - 9]) became evident. A transition to a Mott insulator
becomes possible when a superfluid like a BEC is placed in a periodic
potential. This quantum phase transition is characterized by the competition
between two interactions: the tendency of the particles to hop into adjacted
wells ( kinetic energy ), and the interparticle forces that keep them in
separate wells ( potential energy ). As the strength of the interaction term
(the repulsive interaction energy $ U $ between two atoms ) relative to the
tunneling term (the tunneling $ J $ between adjacent sites  in the Bose - Hubbard
Hamiltonian) is changed, the system reaches a quantum critical point in the
ratio of $ U/J $,
 for which the system will undergo a quantum phase transition from
the superfluid state ( SF ) to the Mott insulator state (MI). A seminal
experiment by Greiner and collaborators [ 5 ] demonstrated a quantum phase
transition in a Bose - Einstein condensate from a superfluid state into a Mott
insulator state, by varying the lattice laser intensity as proposed
theoretically in [ 10 ].

The balance between two interactions, which determines a quantum phase
transition can be changed by a time - dependent action, and thereby a quantum
phase transition can be manipulated. In this 
letter, we propose this
basic idea which will be
used for the SF - MI transition in a BEC and for the external Josephson
effect between two BEC's.

We consider a manipulation of the SF - MI transition in a BEC in an amplitude
-modulated optical lattice. The such
optical
 lattice can be created by
interfering pair of amplitude - modulated beams.
In general,
there are the two generic types of the manipulation.
 The effect of the modulation is to modulate the optical lattice
potential so that it becomes a time - dependent optical lattice potential which
the depth of the potential well is
  $$
        V_0(t) = V_0(l + 
        \varepsilon \cos\omega_Mt), \varepsilon < 1,   \eqno (1)
  $$

where $ \varepsilon $  is the modulation strength, $ \omega_M $ is the
modulation frequency. If the period of modulation $ 2\pi/\omega_M $ is
larger the nonequilibrium dynamic time $\tau_{noneq} $, then the time
behavior of a BEC in an amplitude - modulated optical lattice is quasistatic.
The
nonequilibrium dynamic time $ \tau_{noneq} $ for trapped bosonic atoms in an optical
lattice potential was considered in Ref.[11] and was shown that the
characteristic time of the dynamical restoration
 of the phase coherence in the domain of
a quantum phase transition
is the
integer of the order of 10 multiplied by a Josephson time $ h/J $.
One is a resonant mechanism which operates when the modulation 
 frequency $ \omega_M $ is close to the excited internal state of a BEC.
A resonant mechanism is determined by change in the population due to
a resonant driving.
In a BEC in a periodic potential the characteristic energetic scale of the band
structure [12,13] 
is the recoil energy $ E_R $.
 And, usually, $ E_R $ exceeds $ J $, and the
quasistatic behavior is not valid. However, in this case
is more
convenient to use a picture that an amplitude - modulated field can be
presented as a set of monochromatic fields (main on frequency $ \omega $
 and sideband
on frequencies $ \omega \pm \omega_M $ ). These fields can induce a Raman transition between the two
internal states of a BEC $ |a > $ and $ |b> $.
A
resonant field via a two - photon ( Raman ) transition changes the
population of the ground state of a BEC. Therefore, the on - site interaction
and the tunneling, which depend on the population, are also changed. However,
due to this changing population the variations of the on - site interaction and
the tunneling are different; therefore, the balance between kinetic and
potential energies, which determines a phase transition, is changed. As a
result, we have a modification of the quantum phase transition. 
The modificated by a resonant modulation ratio $ U_r/J_r $ is less than 
this
ratio without a modulation.
The phase transition [6,9,14] occurs at $ U/J = z \times 5.8 = const $, with 
 $ z $ being the number of next neighbours of a lattice site. The
 system
 has the renormalized
  by
   a modulation parameters $ U_r $ and $ J_r $ of a Bose - Hubbard Hamiltonian
 and  reaches a quantum critical point
 if $  {U_r/J_r} = const $
 with the same constant, 
 then it means that the 
 optical potential critical
 depth  in an amplitude - modulated resonant lattice
 $ V_{0{res}}^{cr} $
 increases, because the ratio of $ U/J $ is the growing function [ 10, 14 ]
 on the
 optical potential depth. 
Second is a nonresonant
 mechanism for which we will use a direct time - dependent description.
A nonresonant mechanism of manipulation
 is in operation for a quasistatic regime 
$ \omega_M < 2 \pi \tau_{noneq}^{-1} $, 
therefore, if $ V_0 $ corresponds to a quantum critical point $ V_0^{cr} $
 then as the time
runs, the system goes from a superfluid to an insulator and back again with
frequency $\omega_M $.
The ratio of $ U/J $
is a nonlinear function on the laser intensity with the leading exponential 
dependence on the strength of the periodic potential
$ V_0 $. If we have a time - dependent lattice laser intensity, then,
although $ \overline {V_0(t)} = V_0 $, due to 
the nonlinearity
a time - averaged ratio of  $ \overline{U/J} $ will be different from a 
ratio of
$ U/J $ without a modulation. For a small $ \varepsilon $ we have
 the time - averaged
$ \overline{U/J} = U/J + \Delta(U/J) $, where
the bar stands for a time average
and
$ \Delta(U/J) $
 is proportional to the time - averaged
$ \varepsilon^2 \overline{\cos^2\omega_M t} $ ( the time - averaged of the
modulation harmonic oscillation to the lowest
 even order power ).

{\it A Bose - Einstein condensate in an amplitude - modulate optical lattice.}

 In resonant  case the system is excited
by an off - resonant Raman ( two - photon ) driving with the Rabi frequency
$ \Omega_R $. In an amplitude - modulated
optical lattice two standing laser waves of frequencies $ \omega $
 and $ \omega - \omega_M (—\omega_M \ll \omega ) $ 
 drive an atom in a Raman scheme.
  The  effective two -
photon Rabi frequency is
 given in terms of the single - photon Rabi frequencies $ \Omega $ and 
 $\Omega^{\prime} $
 of the fields on the main $ \omega $ and sideband $\omega^{\prime} =
\omega - \omega_M $ 
frequencies as  
  $ \Omega_R = \frac{\Omega \cdot \Omega^{\prime}}{ 2 \Delta} $.
 For the amplitude - modulated laser beams with
the modulation strength $ \varepsilon $ we have for standing waves
$ \Omega_R({\bf x}) = \frac{ \varepsilon \Omega^2({\bf x})}{ 2 \Delta} =
\frac{ \varepsilon V_0({\bf x}) }{ h } $,
here $ \Delta $ is the far - detuning of the laser beams from the atomic 
resonance 
(the closest neighboring optical dipole transition), and
   $ V_0({\bf x}) =\sum_{j=1}^{3} V_{jo} \cos^2 k x_j $ 
 is the optical lattice potential
 with
  wave vectors  $ k = \frac{ 2 \pi }{ \lambda} $ and $ \lambda $  the wavelength of 
   the laser beams    
 on the main frequency, and we are omitted
the terms proportional to the small parameter
$ \Delta k \cdot x = \frac{\omega_M}{ c } x \ll 1 $,
  where $ \Delta k \equiv k^{\prime} - k $, 
 for $ x < \lambda \sim 10^{-4} cm, \omega_M \preceq 10^7 s^{-1} $,
   this parameter is about
$ 10^{-7} $ . The eigenstates of a BEC are Bloch wave functions, and an appropriate
superposition of Bloch states yields a set of Wannier functions which are
localized on the individual lattice sites. We will consider as the two distinct
internal states $ a $ and $ b $
 coupled via a two - photon ( Raman ) transition the two
different vibrational states 
in the Wannier basis, where the initial state $ a $ is
the lowest vibrational state of the harmonic oscillator with $ n = 0 $
and the
state $ b $ is the excited vibrational state with
$ n \neq 0 $. The equations of motion of
a strongly ( the Rabi regime ) driven [ 15,16 ] BEG resemble the Bloch
equations describing a driven two - level system if the external driving field
$ h \Omega_R $ is much larger
 than the difference in the two - body interaction $ K $ ( this
 energy [17] is proportional to $ U_a + U_b - 2U_{ab} $ ). We consider a steady
state driving regime, in which the solution $ \varphi ({\bf x}) $
of the Bloch equation of
the standard Rabi problem is determined on the parameters of the field ( the
Rabi frequency  and the detuning from resonance 
 $ \delta = \omega_M - ( \omega_b - \omega_a ) $ ). Therefore, the boson
field operator for atom $ \Psi({\bf x}) $ can be represent by an expansion in the Wannier
basis, which is a sum over the lattice $ i $, and multiplied by the function
$ \varphi({\bf x}) $ , which
is determined by the Raman field, 
$ \Psi_a({\bf x}) = \sum_{ i } a_i w_0({\bf x} - {\bf x}_i) \varphi({\bf x}) $,
where the
operators $ a_i $
 are the operators on lattice site $ i $.
 In order to
   avoid a sufficient population of the excited state $ b $, we are assume
 $ \Omega_R < \delta $, then
$ \varphi^2({\bf x}) \simeq 1 - 
\frac{1}{2} (\frac{\Omega_R({\bf x})}{\delta})^2 $.
 In the tight - binding limit the
Wannier states are localized and can be approximated [ 10,14 ] by the harmonic
oscillator wave function
$ w_n({\bf x} - {\bf x}_i) $ of number $ n $.
The size of the ground - state oscillator wave function
$ \beta $ in a lattice is much less than the lattice period
$ \lambda/2, \beta \ll \lambda, $ where the size
 $ \beta = \sqrt{\frac{ h }{ 2 m \nu_j}} $ 
 determined by the atomic
mass $ m $ and the oscillation frequency in the wells ( see, for example [ 10 ] )
$ \nu_j =\sqrt{ 4 E_R V_{jo} }/{\hbar} $, 
 then the inequality
 $ h \Omega_R \gg K $ reduces to the following
 $ \varepsilon V_0 \gg K $.
   This inequality can be
easily fulfiled, because, usually, $  K/h  \sim 1 \div 0.1 s^{-1} $ ( for
 example, $ K/h \simeq 0.1 s^{-1} $
for hyperfine states of
 $ ^{87}Rb $ [18]) and $ V_0 \simeq 10 E_R, E_R/h \simeq 2 \times 10^5 s^{-1} $
 for a blue detuning [ 19 ] $ \lambda = 514nm $ for sodium. 
  We find for 
$ (\frac{\varepsilon V_0}{ h \delta})^2 \ll 1 $
 that the parameters of a Bose - Hubbard Hamiltonian modificated
 by a resonant modulation are
$$
J_r \simeq J [ 1 + \frac{3}{2} (\frac{\varepsilon V_0}{ h \delta})^2],
U_r \simeq U [ 1 - (\frac{\varepsilon V_0}{ h \delta})^2].  \eqno (2)
$$ 
The parameters of a Bose - Hubbard Hamiltonian
 $ J $ and $ U $ 
 depend on the optical lattice depth $ V_0 $, and they were calculated
analytically in Ref.[14] in the tight - binding limit in terms of the
microscopic parameters of the atoms in the optical lattice.
We find that for a resonant
amplitude - modulated optical lattice the ratio
$ U_r/J_r $ is always less than without a modulation,
then it means that for a resonant modulation the optical potential critical
depth $ V_0^{cr} $ increases.
  The corresponding tuning  $ \Delta V_0^{res}(\varepsilon) $
of the optical lattice critical depth
 is determined for $ V_0 \gg E_R $  by 
 
 $$ 
   \Delta V_0^{res}(\varepsilon) \equiv \frac{ V_{0{res}}^{cr}(\varepsilon) -
   V_0^{cr} }{ V_0^{cr}} \simeq \frac{5}{2} 
    (\frac{\varepsilon V_0^{cr}}{ h \delta})^2
   (\frac{E_R}{V_0^{cr}})^{1/2},              \eqno (3)
   $$

As a result, we can manipulate the superfluid - Mott insulator transition of a
BEC in an amplitude - modulated optical lattice. The tuning of the optical
lattice critical depth is determined by
the modulation 
strength $ \varepsilon $.
 By varying the parameter  $ \varepsilon / \delta $,
 we can tune the critical value to a given value.
Therefore, in the interval of the laser intensity $ V_0 $, for which 
$  V_0^{cr}  <  V_0  <  V_{0 {res}}^{cr}(\varepsilon)   $, 
 we can switch one phase to another by a 
modulation, and in this interval of the
initial 
 MI  phase  a modulation 
switches the  MI  phase to the  SF phase. 
 
 For
 a nonresonant manipulation we will use
  a
time - dependent potential ( Eq.(l)), and there could be a modulation
frequency $ \omega_M $
 low enough for quasistatic time behavior. Therefore, it modulates
the on - site two - body interaction $ U $
and the hopping matrix element $ J $, so
that these quantities become time - dependent $ U(t),J(t) $. Note that such
potential is created by interfering pair of amplitude - modulated laser beams
with the modulation strength $ \varepsilon/2 $
of the electric field amplitude, and was
omitted the quadratic term for the small parameter $ \varepsilon^2/4  $.
Using the analytical dependencies [ 14, 20 ], we have 
for a time - dependent optical
lattice depth
   $$
  U(t) =\sqrt{2 \pi} (\frac{g}{h}) (\frac{ V_0(t)}{E_R})^{1/4},
  J(t) = \frac{4}{\sqrt{\pi}}E_R (\frac{ V_0(t)}{E_R})^{3/4} 
  exp[-2(\frac{V_0(t)}{E_R})^{1/2}].        \eqno (4)
  $$
 Calculating the time - average of the product of $ U(t) $ by $ (J(t))^{-1} $
 by expanding in a series about the small parameter
 $ \varepsilon^2 \frac{V_0}{E_R} < 1 $,
we find
 $$
 \frac{ (\overline{\frac{U}{J}})_{nr} - 
 (\frac{ U }{ J }) }{ (\frac{U}{J}) } \simeq \frac{\varepsilon^2}{4}
 [\frac{V_0}{E_R} - (\frac{V_0}{E_R})^{1/2} - \frac{1}{2}],     \eqno (5) 
  $$
where $ g = 2h \omega a_s , \omega $ is the confining frequency,
$ a_s $ is the $ s $ -wave scattering length between two atoms,
 $ (\overline{\frac{U}{J}})_{nr} $
  is the time - average of the ratio $ \frac{ U }{ J } $ for a nonresonant
  modulation, and
   were omitted the small terms
 proportional to $ (\varepsilon^2 \frac{V_0}{E_R})^n $ for $ n > 1 $.
If $ \varepsilon^2 \frac{ V_0}{E_R} $ is not small, then
 we may expand the expression for the time - dependent $ J(t) $ using Bessel
 functions $ I_n( x ) $ and the functions $ \cos^n \omega_M t $.
 In this case instead the parameter $ \frac{\varepsilon^2}{4} 
 [\frac{V_0^{cr}}{E_R} - (\frac{V_0^{cr}}{E_R})^{1/2} - \frac{1}{2}] $ we have
 $ I_2( \varepsilon \sqrt{\frac{V_0^{cr}}{E_R}}) $, because, usually,
 $ V_0/E_R \sim 10 $ and for $ \varepsilon < 0.4 $ among $ I_n( x < 1.3 ) $
 the largest is  $ I_2(x) $.

{\it An external Josephson effect between two BECs.}

Here we consider briefly the manipulation
by an amplitude - modulated laser beam
  an external Josephson effect
between two BECs, the full presentation will be given elsewhere.

 This system is a prototype system of an array of
Josephson junctions. In such systems, a superconducting - insulator phase
transition is driven by the competition between the Josephson coupling energy
$ E_j $, which govern the tunneling through the intrawell barriers, and the on -
site interparticle interaction energy $ E_C $.
A Josephson effect between two spatially separate BECs can be realized only
with small barriers. A key element of this effect is the Josephson coupling
energy $ E_j $. Based on a tunneling Hamiltonian description the dominant
Josephson term of first order in the tunneling is proportional to the barrier
transmission amplitude. Such approximation by a tunneling or transfer
Hamiltonian is valid for high and narrow barriers, for which the barrier
transmission amplitude $ t_{lr} $ is proportional to
$ exp(-k_{\mu} d) $, where $ k_{\mu} = \sqrt{2m( V_B -
 \mu )/h^2}, \mu = (\mu_a - \mu_b)/2 $
 is the average chemical potential, where the two BECs
( $ a $ and $ b $) confined in a cubic volume $ L^3 $
are separated by a square potential
barrier of height $ V_B $ and width $ d $. A Josephson geometry for BECs may be created
by splitting a single condensate in a long trap into two separate parts by a
narrow light sheet produced by a strongly detuned laser beam. Because the
repulsive potential due to the ac Stark shift is proportional to the laser
power, thus the height of the barrier is determined by the laser intensity.
Therefore, if we create the necessary for a Josephson effect two - well
potential by an amplitude - modulated laser beam, then the potential barrier
height $ V_B $ will be
 an amplitude - modulated time - dependent one ( similarly Eq.(1)).
  Just as the analytical dependencies of the tunneling $ J $ 
 between
adjacent sites of a BEC in an amplitude - modulated optical lattice and the
transmission amplitude $ t_{lr} $
 are similar, so the time - average of this barrier
transmission amplitude $ \overline{ t }_{lr} $
 is differ from the value $ t_{lr} $ without a modulation.
As a result, the Josephson coupling energy $ E_j $, which is proportional to the
barrier transmission amplitude, is also can be manipulated
by the abovementioned manner.

In summary, the superfluid - Mott insulator transition in a BEC
in an optical lattice 
can be manipulated 
by an amplitude - modulated manner. 
This manipulating is caused by the change of the balance  two interactions,
which determines a quantum phase transition, by a time - dependent action.
There are two 
 mechanisms of the manipulation.
  One, resonant mechanism is in operation for the modulation frequency
 closed to the excited internal state of a BEC, and
 two standing laser waves of main and sideband frequencies of an
 amplitude - modulated optical lattice induce a Raman transition.
 Due to resonant Raman driving
   the population of the ground state of a BEC changes. 
 Therefore, the on - site interaction and the atomic tunneling 
 are also changed, and their ratio is different from the ratio without a
 resonant modulation. As a result, the
 optical lattice critical depth is other than without a modulation, this
 tuning is
  determined by the square of the modulation strength.
  The optical lattice critical depth with a resonant modulation is higher than
  without a modulation, and, therefore, there is an interval of the initial
  MI phase, in which a modulation switches the MI phase to the SF phase.
  Second, nonresonant mechanism is determined by the nonlinear dependence
  of the ratio
 of the repulsive interaction to the tunneling on the optical lattice depth.
 Due to this nonlinearity for the harmonic modulation of the optical lattice
 depth the time - averaged ratio is different from the ratio without a
 modulation.
  The proposed
mechanism of manipulating and controlling Bose - Einstein condensates makes it
possible to tune the critical value of parameter of the quantum phase
transition to a given value.

In conclusion, tuning and conducting, in general, the superfluid - Mott
insulator transition in a BEC in an optical lattice by 
a time - dependent action ( in particular, in an amplitude - modulated optical
lattice )
 can be a path to the solution of the general problem of controlling a
quantum phase transition.

I thank A.J.Freeman for encouragement, A. Patashinski for discussions.

\end{document}